\documentclass[11pt,a4paper]{article}

\usepackage{amsmath,amsfonts,graphicx,natbib,psfrag,color}
\usepackage{algorithm, algorithmic}
\usepackage[textsize=small]{todonotes}
\usepackage{booktabs}
\usepackage[capitalise, noabbrev]{cleveref}
\usepackage[parfill]{parskip}
\usepackage{multirow}
\usepackage{mathtools}

\title{Sequential Bayesian inference for stochastic epidemic models of cumulative incidence}

\author{Sam A. Whitaker$^1$\footnote{s.whitaker2@newcastle.ac.uk}, Andrew Golightly$^2$\footnote{andrew.golightly@newcastle.ac.uk},
Colin S. Gillespie$^1$\footnote{colin.gillespie@newcastle.ac.uk}, Theodore Kypraios$^3$\footnote{theodore.kypraios@nottingham.ac.uk}}
\date{\small $^1$ School of Mathematics, Statistics and Physics, Newcastle University, UK\\$^2$ Department of Mathematical Sciences, Durham University, UK\\$^3$ School of Mathematical Sciences, University of Nottingham, UK}
\begin{document}
\maketitle
\begin{abstract}
\noindent Epidemics are inherently stochastic, and stochastic models provide an appropriate way to describe and analyse such phenomena. Given temporal incidence data consisting of, for example, the number of new infections or removals in a given time window, a continuous-time discrete-valued Markov process provides a natural description of the dynamics of each model component, typically taken to be the number of susceptible, exposed, infected or removed individuals. Fitting the SEIR model to time-course data is a challenging problem due incomplete observations and, consequently, the intractability of the observed data likelihood. Whilst sampling based inference schemes such as Markov chain Monte Carlo are routinely applied, their computational cost typically restricts analysis to data sets of no more than a few thousand infective cases. Instead, we develop a sequential inference scheme that makes use of a computationally cheap approximation of the most natural Markov process model. Crucially, the resulting model allows a tractable conditional parameter posterior which can be summarised in terms of a set of low dimensional statistics. This is used to rejuvenate parameter samples in conjunction with a novel bridge construct for propagating state trajectories conditional on the next observation of cumulative incidence. The resulting inference framework also allows for stochastic infection and reporting rates. We illustrate our approach using synthetic and real data applications.
\end{abstract}


\section{Introduction}
\label{sec:intro}
Statistical epidemiology is a large and ever-growing field that combines statistical models with state-of-the-art inference techniques to analyse complex data arising from observational studies. Epidemics are inherently random, and capturing this aspect necessitates the use of stochastic models, which typically take the form of a continuous-time, discrete-valued Markov jump process (MJP), describing transitions between different states such as susceptible, exposed, infected, removed \citep[SEIR, see e.g.][]{allen17}. Inference for the parameters governing such models is complicated by the availability of incomplete and imperfect observation regimes, as arising, for example, when only a subset of states or transitions are recorded at discrete times. This problem can be circumvented via the use of Markov chain Monte Carlo and data augmentation \citep[see e.g.][]{oneill99,jewell09} although these schemes appear limited to applications where the total population size is of the order of a few thousand individuals \citep{stockdale19}. Therefore, a key challenge is the development of fast and reliable inference techniques to allow real-time decision making \citep{SWALLOW22}.

Computationally efficient approaches to inference for stochastic epidemic models include the use of approximate inference schemes \citep[see e.g.][for approximate Bayesian computation schemes]{KYPRAIOS17,McKinley18,MINTER19} or direct approximation of the most natural MJP \citep[see e.g.][for approximations based on stochastic differential equations]{cauchemez08,fuchs2013inference,fintzi2021linear} and direct approximation of the observed data likelihood \citep[see e.g.][]{whiteley2021, golightly23}. In this paper, we replace the MJP model of cumulative incidence with a model in which the number of transition events over a time interval whose length is chosen by the practitioner is assumed to be Poisson distributed. The resulting time discretised SEIR model (dSEIR) has a number of advantages over other approximate models, including recognition of the discrete stochastic nature of epidemic spread, easy incorporation of time varying parameters, specification of a time step that trades off accuracy and computational efficiency and, crucially, tractability of the conditional posterior for rate parameters, given a particular choice of prior.

We fit the dSEIR model within a sequential Bayesian framework. That is, we update our inferences about the static parameters and latent dynamic process as each observation becomes available. The observations are noisy measurements of cumulative incidence, specifically, the number of new infections or removals in fixed-length time windows. The inference procedure uses sequential Monte Carlo \citep[see e.g.][]{KANTAS09,chopin2020introduction} via recursive application of a series of propagate, weight, resample and rejuvenate steps. The latter is used to lessen sample impoverishment of the static parameter particle set and leverages the tractability of the parameter posterior conditional on the latent dynamic process, summarised through a collection of summary statistics. This approach was first described in \cite{storvik2002} \citep[see also][]{fearnhead2002} and is a key ingredient of particle learning \citep{carvalho2010}, which has been used by \cite{dukic12} and \cite{lin2013b}, among others, in the context of stochastic epidemic models. Our approach is most related to the latter, which considers a particle learning algorithm for the most natural MJP representation of an epidemic compartment model of prevalence data, coupled with a latent seasonal component. However, unlike their approach, our modelling framework allows for additional flexibility via the use of a stochastic differential equation (SDE) to describe time varying parameters (e.g. infection and/or reporting rates) and the specification of an observation model to allow for imperfect observations on cumulative incidence. Moreover, we adapt the bridge construct of \cite{golightly2015} \citep[see also][]{GoliKyp17} to the context of cumulative incidence, so that particle trajectories are propagated conditional on the following observation. We illustrate the resulting inference scheme in three scenarios: the first uses synthetic data in order to assess the accuracy of our approach for different numbers of particles (bench-marked against the output of a pseudo-marginal Metropolis-Hastings scheme \citep[PMMH, e.g.][]{andrieu2010particle} while the two final scenarios consider the spread of Ebola in West Africa \citep{fintzi2021linear} and COVID-19 in New York \citep{aspannaus2021bayesian}.       

The remainder of this paper is organised as follows. Section~\ref{sec:sem} describes the Markov jump process representation of cumulative incidence in an SEIR setting before considering the time discretised approximation. Section \ref{sec:sbinf} outlines our sequential approach to Bayesian inference and gives the bridge construct. Applications are considered in Section~\ref{sec:applications}, and conclusions are drawn in Section~\ref{sec:sum}. 

\section{Stochastic epidemic models}
\label{sec:sem}
We consider SIR \citep{AnBr00,kermack27} and SEIR models \citep{hethcote2000} within which a population of fixed size $N_{\textrm{pop}}$ is classified into compartments consisting of susceptible ($S$), exposed ($E$), infectious ($I$) and removed ($R$) individuals. In what follows, we describe the stochastic SEIR model as the most general case, and make clear how the SIR model can be obtained as a simplification thereof.

Transitions between compartments can be described by the set of pseudo-reactions given by
\[
S+I \xrightarrow{\beta} E +I, \quad E \xrightarrow{\kappa} I, \quad I \xrightarrow{\gamma} R . 
\]
The first transition describes contact of an infective individual with a susceptible and with the net effect resulting in an exposed individual and one fewer susceptible. The second transition moves an exposed individual to the infected class, and the final transition accounts for an infected individual's removal (recovered with immunity, quarantined or dead). The components of $\theta=(\beta,\kappa,\gamma)'$ denote contact, infection and removal rates. Note that  setting $\kappa=0$ and substituting $E$ for $I$ gives the set of pseudo-reactions governing the SIR model as \[
S+I \xrightarrow{\beta} 2I, \quad I \xrightarrow{\gamma} R . \]
Let $X_t=(S_t,E_t,I_t)'$ denote the numbers in each state at time $t\geq 0$ and note that $R_t=N_{\textrm{pop}}-S_t-E_t-I_t$ for all $t\geq 0$. The dynamics of $\{X_t,t\geq 0\}$ are most naturally described by a Markov jump process (MJP), that is, a continuous time, discrete valued Markov process. Assuming that at most one event can occur over a time interval $(t,t+\Delta t]$ and that the state of the system at time $t$ is $x_t=(s_t,e_t,i_t)'$, the MJP is characterised by transition probabilities of the form 
\begin{align*}
\mathbb{P}(X_{t+\Delta t}=(s_t-1,e_t+1,i_t)'|x_t,\theta) &= \beta s_t i_t \,\Delta t+o(\Delta t),\\
\mathbb{P}(X_{t+\Delta t}=(s_t,e_t-1,i_t+1)'|x_t,\theta) &= \kappa e_t \,\Delta t+o(\Delta t),\\
\mathbb{P}(X_{t+\Delta t}=(s_t,e_t,i_t-1)'|x_t,\theta) &= \gamma i_t \,\Delta t+o(\Delta t),\\
\mathbb{P}(X_{t+\Delta t}=(s_t,e_t,i_t)'|x_t,\theta) &= 1-(\beta s_t i_t+\kappa e_t + \gamma i_t)\,\Delta t+o(\Delta t),
\end{align*}
and $o(\Delta t)/\Delta t\to 0$ as $\Delta t\to 0$. Similarly, the cumulative incidence of contact, infection and removal events $\{N_t, t\geq 0\}$ is an MJP governed by the transition probabilities 
\begin{align*}
\mathbb{P}(N_{t+\Delta t}=(n_{se}+1,n_{ei},n_{ir})'|n_t,x_t,\theta) &= \beta s_t i_t \,\Delta t+o(\Delta t),\\
\mathbb{P}(N_{t+\Delta t}=(n_{se},n_{ei}+1,n_{ir})'|n_t,x_t,\theta) &= \kappa e_t \,\Delta t+o(\Delta t),\\
\mathbb{P}(N_{t+\Delta t}=(n_{se},n_{ei},n_{ir}+1)'|n_t,x_t,\theta) &= \gamma i_t \,\Delta t+o(\Delta t),\\
\mathbb{P}(N_{t+\Delta t}=(n_{se},n_{ei},n_{ir})'|n_t,x_t,\theta) &= 1-(\beta s_t i_t+\kappa e_t + \gamma i_t)\,\Delta t+o(\Delta t),
\end{align*}
where $n_t=(n_{se},n_{ei},n_{ir})'$ denotes the contact, infection and removal events. 

It will be helpful to define the instantaneous rate or hazard function $h(x_t)=(h_1(x_t),h_2(x_t),\linebreak h_3(x_t))'$ by 
\begin{align*}
h(x_t)&=\lim_{\Delta t\to 0} \mathbb{P}(X_{t+\Delta t}|x_t,\theta) / \Delta t\\
&= (\beta s_t i_t,\kappa e_t, \gamma i_t)'
\end{align*}
and we suppress the dependence of the hazard function on $\theta$ to simplify the notation. Additionally, we define the net effect matrix by
\[
A=\begin{pmatrix}
-1           & \phantom{-}1 & \phantom{-}0\\
\phantom{-}0 & -1           & \phantom{-}1\\
\phantom{-}0 & \phantom{-}0 & -1
\end{pmatrix}
\]
where it is understood that each row describes the effect on each component of $X_t=(S_t, E_t, I_t)'$ by the respective occurrence of a contact, infection or removal event. Then, the processes describing the prevalence $\{X_t,t\geq 0\}$ and cumulative incidence $\{N_t, t\geq 0\}$ are linked via the equation
\[
X_t = x_0 + \sum_i A_i' N_{i,t}
\]
where $x_0$ represents the initial number of susceptible, exposed and infected individuals, $A_i$ denotes the $i$th row of $A$ and $N_{i,t}$ denotes the $i$th component of the incidence process at time $t$.

Given $x_0$ and $\theta$, generating exact realisations of the incidence process, and therefore the stochastic SEIR model is straightforward and can be achieved by using well-known simulation algorithms from the stochastic kinetic model's literature \citep[see, e.g.][]{wilkinson2018}. The simplest approach is Gillespie's direct method \citep{gillespie1977exact}, which simulates the time to the next event as an exponential random variable with rate $\sum_{i=1}^3 h_i$. The event that occurs will be of type $i$ (with $1=$ exposure, $2=$ infection, $3=$ removal) with probability proportional to $h_i$. 

\subsection{Time varying contact rate \label{subsec:tvcr}}

In practice, assuming that the contact rate in the SEIR model (or infection rate in the SIR model) remains constant throughout the epidemic may be unreasonable. We therefore follow \cite{dureau13,aspannaus2021bayesian,wadkin22} among others and describe the contact rate via an It\^o stochastic differential equation (SDE). Let $\{\beta_t,t\geq 0\}$ denote the infection process and consider $\tilde{\beta}_t = \log(\beta_t)$, assumed to satisfy a time-homogeneous SDE, parameterised by $\lambda$ and of the form
\begin{equation}
        \text{d}\tilde{\beta}_t = a(\tilde{\beta}_t,\lambda)
        \text{d}t + 
        b(\tilde{\beta}_t,\lambda)\text{d}W_t
        \label{eqn:TVB}
\end{equation}
where $\{W_t,t \geq 0\}$ is a standard Brownian motion process. 
Choice of the drift and diffusion functions $a(\cdot)$ and $b(\cdot)$ in \cref{eqn:TVB} determine the properties of  $\tilde{\beta}_t$. For example, $a(\tilde{\beta}_t,\lambda)=0$ and $b(\tilde{\beta}_t,\lambda)=\lambda$ gives a generalised Brownian motion process (as used in \cref{subsec:sim_study,subsec:covid}) which admits an analytic solution over a time interval $(t,t+\Delta t]$ as
\begin{equation}\label{sde}
\tilde{\beta}_{t+\Delta t}=\tilde{\beta}_t +\lambda^{-1/2} \Delta W_t
\end{equation}
where $\Delta W_t:=(W_{t+\Delta t}-W_t)\sim \textrm{N}(0,\Delta t)$ and $\lambda$ is a precision parameter. In cases where the SDE (\ref{eqn:TVB}) can't be solved analytically, a numerical approximation can be sought. The simplest such approximation is given by the Euler-Maruyama discretisation
\begin{equation}\label{EM}
\tilde{\beta}_{t+\Delta t}=\tilde{\beta}_t + a(\tilde{\beta}_t,\lambda)
        \Delta t + 
        b(\tilde{\beta}_t,\lambda)\Delta W_t .
\end{equation}
Replacing $\beta$ with a time-varying $\beta_t=\exp(\tilde{\beta}_t)$ dictates that the hazard $h_1$ of the contact reaction is no longer constant between event occurrences. Generating exact realisations from the resulting SEIR model is no longer straightforward unless $\beta_t$ can be bounded above, permitting the use of Poisson thinning \citep{lewis1979}. In what follows, we consider a time discretised stochastic SEIR model, and subsequently base our inferential approach on the resulting approximation. In particular, the practitioner can choose the discretisation level to balance accuracy and computational efficiency.

\subsection{Time discretised model}

Consider a time interval $(t,t+\Delta t]$ and denote by $\Delta N_{t}=(\Delta N_{1,t},\Delta N_{2,t},\Delta N_{3,t})'$ the length-3 vector containing the number of events of each type (exposure, infection, removal) over this interval. We assume that $\Delta t$ is small enough to reasonably assume that $\tilde{\beta}_s\approx \tilde{\beta}_t$ and $X_s\approx x_t$ for $s\in (t,t+\Delta t)$. It should then be clear that the $i$th component of $\Delta N_{t}$ follows a Poisson distribution with rate $h_{i}(x_t,\tilde{\beta}_t)\Delta t$, where we now let the hazard function depend additionally on $\tilde{\beta}_t$. Explicitly,
\[
\Delta N_{1,t}\sim \textrm{Po}(\exp(\tilde{\beta}_t)\, s_t i_t\Delta t), \quad 
\Delta N_{2,t}\sim \textrm{Po}(\kappa \,e_t\Delta t), \quad
\Delta N_{3,t}\sim \textrm{Po}(\gamma \,i_t\Delta t).
\]
Hence, we obtain the time discretised SEIR model (dSEIR) with time-varying contact rates as
\begin{equation}\label{dSEIR}
\left\{ \begin{array}{l}
X_{t+\Delta t}= x_{t} + A' \Delta N_{t},\\
\tilde{\beta}_{t+\Delta t}=\tilde{\beta}_t + a(\tilde{\beta}_t,\lambda)
        \Delta t + 
        b(\tilde{\beta}_t,\lambda)\Delta W_t .
\end{array}
\right.
\end{equation}
Simulating exact realisations from dSEIR is now straightforward; see \cref{alg:PoL} for further details.
\begin{algorithm}[t]
        \caption{dSEIR: Forward Simulation}
        \label{alg:PoL}
        \textbf{Input:} parameters $\theta=(\kappa,\gamma)'$, infection process parameters $\lambda$, initial conditions $x_0=(s_0,e_0,i_0)'$ and $\tilde{\beta}_0$, time step $\Delta t$ and end time $T=m\Delta t$.\\
For $j=0,\ldots,m-1$:
        \begin{enumerate}
\item Set $t:=j\Delta t$ and calculate the hazard function $h(x_t,\tilde{\beta}_t)=(\exp(\tilde{\beta}_t)\, s_t i_t,\kappa \,e_t,\gamma \,i_t)'$.
\item Simulate the incidence increment $\Delta N_t$ by drawing $\Delta N_{i,t}\sim \textrm{Po}(h_i\Delta t)$, $i=1,2,3$. Set $x_{t+\Delta t}= x_{t} + A' \Delta N_{t}$.
\item Simulate $\Delta W_t\sim \textrm{N}(0,\Delta t)$ and set $\tilde{\beta}_{t+\Delta t}=\tilde{\beta}_t + a(\tilde{\beta}_t,\lambda)
        \Delta t + 
        b(\tilde{\beta}_t,\lambda)\Delta W_t$
        \end{enumerate}
\textbf{Output:} trajectory $\{(x_t,\tilde{\beta}_t), t=0,\Delta t,\ldots,T\}$.      
\end{algorithm}

\section{Sequential Bayesian inference}
\label{sec:sbinf}

In this section, we consider the problem of performing full Bayesian inference for the parameters (and unobserved dynamic processes) governing the dSEIR model based on incidence observations that may be incomplete and subject to measurement error. We describe the observation model before considering the inference task. We perform sequential Bayesian inference via a particle filter, a key ingredient of which is a novel construct that allows approximate draws of the dSEIR process between observation instants. 

\subsection{Observation model}
\label{sec:obsmodel}
Consider data $y=(y_{t_1},\ldots, y_{t_L})'$ on a regular time grid, where $y_{t_i}$ is a (partial) observation on the cumulative incidence $\Delta N_{t_i}=N_{t_{i}}-N_{t_{i-1}}$ over a time interval $(t_{i-1},t_i]$. We assume that the incidence process is not observed directly and consider two observation models as follows. The first model reflects the assumption that cases are under-reported, leading to
\begin{equation}\label{eqn:obseqn1}
	Y_{t_i}|\Delta N_{t_i} \sim \textrm{Bin}(P'\Delta N_{t_i},\rho), \quad i=1,\ldots, L.
\end{equation}
Here, $P$ is a constant matrix allowing for observation of a subset of components of $\Delta N_{t_i}$ and $\rho$ controls the accuracy of the observation process. In practice, we take $P'=(0,1,0)$ in the SEIR model and $P'=(1,0)$ in the case of the SIR model so that observations are noisy counts of new infections in a given time window.  The second observation model allows for over-dispersion in the observation process via
\begin{equation}\label{eqn:obseqn2}
	Y_{t_i}|\Delta N_{t_i} \sim \textrm{NegBin}(\mu_i= \rho P'\Delta N_{t_i},\sigma^2_i=\mu_i+\mu_i^2/\nu), \quad i=1,\ldots, L
\end{equation}
where $\textrm{NegBin}(\mu,\sigma^2)$ denotes a Negative Binomial random variable with mean $\mu$ and variance $\sigma^2)$. For simplicity, we have assumed that both observation models use a constant reporting rate $\rho$. We relax this assumption in Section~\ref{subsec:covid} by considering a dynamic reporting rate, assumed to follow a (discretised) SDE; see Section~\ref{subsec:covid} for further details. 

We denote by $\phi$ the parameters governing the observation model and explicitly link the data to the latent incidence process via the mass function $\pi(y_{t_i}|\Delta n_{t_i},\phi)$.

\subsection{Inference task}
\label{sec:inftask}
We assume that interest lies in the vector of all static parameters $\psi=(\theta',\lambda',\phi')'$, the latent incidence process $\{N_t, t_0\leq t\leq t_L\}$, contact rate process $\{\tilde{\beta}_t, t_0\leq t \leq t_L \}$ and the initial state vector $x_{t_0}$. Note that the initial state vector and incidence process are sufficient to determine the prevalence process $\{X_t, t_0\leq t\leq t_L \}$ deterministically, through recursive application of the first equation in (\ref{dSEIR}). In what follows, we drop explicit dependence of the incidence process on $x_{t_0}$ from the notation for ease of exposition.   

Recall that $y_{t_i}$ denotes a noisy and incomplete observation on the cumulative incidence $\Delta N_{t_i}$ over $(t_{i-1},t_{i}]$. We assume that the inter-observation interval $\Delta t=t_{i}-t_{i-1}$ is too large, over which event hazards can be assumed plausibly constant. We, therefore partition each such time interval as
\[
t_{i-1} = \tau_{i,0}<\tau_{i,1}<\cdots < \tau_{i,m}=t_{i}
\]
with $\tau_{i,j}-\tau_{i,j-1}=\Delta\tau=\Delta t/m$. This allows the practitioner to choose the discretisation level $m$ to balance the accuracy and computational efficiency of the dSEIR model (\ref{dSEIR}). 

We let $\tilde{\beta}=(\tilde{\beta}_{\tau_{1,0}},\tilde{\beta}_{\tau_{1,1}},\ldots,\tilde{\beta}_{\tau_{L,m}})'$ and $\Delta n=(\Delta n_{\tau_{1,1}},\Delta n_{\tau_{1,2}},\ldots,\Delta n_{\tau_{L,m}})'$ denote the latent contact rate process and collection of incidences over sub-intervals $(\tau_{i,j-1},\tau_{i,j}]$, for $i=1,\ldots,L$ and $j=1,\ldots,m$. Note that $\Delta n_{t_i}=\sum_{j=1}^m \Delta n_{\tau_{i,j}}$ then gives the cumulative incidence over $(t_{i-1},t_{i}]$. Upon ascribing a prior density $\pi(\psi)$ to $\psi$, the inference tasks proceeds via the joint posterior
\begin{align}\label{jpost}
\pi(\psi,\tilde{\beta},\Delta n|y)& \propto \pi(\psi) \pi(\tilde{\beta}|\lambda)\pi(\Delta n|\tilde{\beta},\theta)\pi(y|\Delta n,\phi).
\end{align}
Here, the joint density of the latent contact rate process is
\begin{align*}
\pi(\tilde{\beta}|\lambda)&=\pi(\tilde{\beta}_{\tau_{1,0}})\prod_{i=1}^L\prod_{j=1}^m \pi(\tilde{\beta}_{\tau_{i,j}}|\tilde{\beta}_{\tau_{i,j-1}},\lambda)\\
&= \pi(\tilde{\beta}_{\tau_{1,0}})\prod_{i=1}^L\prod_{j=1}^m \textrm{N}(\tilde{\beta}_{\tau_{i,j}}; a(\tilde{\beta}_{\tau_{i,j-1}},\lambda)
        \Delta \tau \,,\, b^2(\tilde{\beta}_{\tau_{i,j-1}},\lambda)\Delta \tau ),
\end{align*}
where $\textrm{N}(\cdot;m,v^2)$ denotes the density of a normal random variable with mean $m$ and variance $v^2$. The joint probability of the incidence process is
\begin{align*}
\pi(\Delta n|\tilde{\beta},\theta)
&= \prod_{i=1}^L\prod_{j=1}^m \prod_{k=1}^3 \textrm{Po}(\Delta n_{k,\tau_{i,j}}; h_{k}(x_{\tau_{i,j-1}},\tilde{\beta}_{\tau_{i,j-1}})\Delta\tau )
\end{align*}
where $\textrm{Po}(\cdot;h)$ denotes the probability mass function of a Poisson random variable with mean $h$. Finally, 
\[
\pi(y|\Delta n,\phi) = \prod_{i=1}^L \pi(y_{t_i}|\Delta n_{t_i},\phi)
\]
with $\pi(y_{t_i}|\Delta n_{t_i},\phi)$ as the probability mass function arising from either (\ref{eqn:obseqn1}) or (\ref{eqn:obseqn2}). 

Since the joint posterior in (\ref{jpost}) will be intractable, we resort to Monte Carlo methods for generating samples of the parameters and latent dynamic processes. In particular, we wish to perform inference sequentially and develop a particle filter approach in the next section.

\subsection{Particle filter approach}\label{sec:pf}

It will be helpful here to introduce the shorthand notation $y_{[1, i]}=(y_{t_1},\ldots,y_{t_i})'$ to denote the observations up to (and including) time $t_i$. Similarly, $\tilde{\beta}_{[0,i]}=(\tilde{\beta}_{\tau_{1,0}},\tilde{\beta}_{\tau_{1,1}},\ldots,\tilde{\beta}_{\tau_{i,m}})'$ and $\Delta n_{(0,i]}=(\Delta n_{\tau_{1,1}},\Delta n_{\tau_{1,2}},\ldots,\Delta n_{\tau_{i,m}})'$ denote respectively, the latent contact rate process and collection of incidence increments over the corresponding time horizon.  

Now, by applying Bayes theorem sequentially, we obtain the posterior  distribution at time $t_{i+1}$ as 
\begin{align}\label{seqpost}
\pi(\psi,\tilde{\beta}_{[0,i+1]},\Delta n_{(0,i+1]}|y_{[1,i+1]}) &\propto 
\pi(\psi,\tilde{\beta}_{[0,i]},\Delta n_{(0,i]}|y_{[1,i]})\nonumber\\
& \quad \times \pi(\tilde{\beta}_{(i,i+1]}|\tilde{\beta}_{t_i},\lambda)\pi(\Delta n_{(i,i+1]}|\tilde{\beta}_{[i,i+1)},\theta)\pi(y_{t_{i+1}}|\Delta n_{t_{i+1}},\phi)
\end{align}
and note that, $\tilde{\beta}_{(i,i+1]}=(\tilde{\beta}_{\tau_{i+1,1}},\ldots,\tilde{\beta}_{\tau_{i+1,m}})'$ whereas 
$\tilde{\beta}_{[i,i+1)}=(\tilde{\beta}_{\tau_{i+1,0}},\ldots,\tilde{\beta}_{\tau_{i+1,m-1}})'$. Now, given an equally weighted sample of `particles'  $\left(\psi^{(1:N)},\tilde{\beta}_{[0,i]}^{(1:N)},\Delta n_{(0,i]}^{(1:N)}\right)$ from $\pi(\psi,\tilde{\beta}_{[0,i]},\Delta n_{(0,i]}|y_{[1,i]})$, the form of (\ref{seqpost}) immediately suggests an importance resampling step that extends the dynamic process particles via  $\tilde{\beta}_{(i,i+1]}^{(k)}\sim \pi(\cdot|\tilde{\beta}_{t_i}^{(k)},\lambda^{(k)})$, $\Delta n_{(i,i+1]}^{(k)}\sim \pi(\cdot|\tilde{\beta}_{[i,i+1)}^{(k)},\theta^{(k)})$, weights the resulting particles $\left(\psi^{(k)},\tilde{\beta}_{[0,i+1]}^{(k)},\Delta n_{(0,i+1]}^{(k)}\right)$ by $w_{i+1}^{(k)}\propto \pi(y_{t_{i+1}}|\Delta n_{t_{i+1}}^{(k)},\phi^{(k)})$, before  sampling with replacement among the particles (using the weights as probabilities). Applying this sequence of steps recursively in time, with samples output at each observation instant (summarising the sequence of filtering distributions $\pi(\psi,\tilde{\beta}_{t_i},\Delta n_{t_i}|y_{[1,i]})$, $i=1,\ldots,L$), gives the bootstrap particle filter \cite[see e.g.][]{gordon1993novel}. Two problems are apparent here. Firstly, repeated resampling of static parameter particles will lead to sample impoverishment (with marginal parameter posteriors collapsing to point masses). Secondly, drawing the incidence process `blindly' (that is, via forward simulation) is likely to lead to many particle trajectories with negligible weight.

To lessen the problem of sample impoverishment, we follow the approach of \cite{storvik2002} (and see also \cite{fearnhead2002}) by noting that the conditional posterior of a subset of parameter components is tractable for a particular choice of prior. To this end, consider the factorisation
\begin{align*}
\pi(\psi,\tilde{\beta}_{[0,i]},\Delta n_{(0,i]}|y_{[1,i]})
&=\pi(\theta,\lambda|\phi,\tilde{\beta}_{[0,i]},\Delta n_{(0,i]},y_{[1,i]})
\pi(\phi,\tilde{\beta}_{[0,i]},\Delta n_{(0,i]}|y_{[1,i]})\\
&=\pi(\theta,\lambda |\Delta n_{(0,i]},\tilde{\beta}_{[0,i]}) \pi(\phi,\tilde{\beta}_{[0,i]},\Delta n_{(0,i]}|y_{[1,i]})\\
&=\pi(\theta |\Delta n_{(0,i]}) \pi(\lambda | \tilde{\beta}_{[0,i]}) \pi(\phi,\tilde{\beta}_{[0,i]},\Delta n_{(0,i]}|y_{[1,i]})
\end{align*}
where the last two lines follow from the conditional independence relationships present in the model. Assuming an independent prior specification for the components of $\theta=(\kappa,\gamma)'$, with $\kappa\sim \textrm{Gamma}(a_\kappa,b_\kappa)$ and $\gamma\sim \textrm{Gamma}(a_\gamma,b_\gamma)$ gives
\begin{equation}\label{cp1}
\kappa|\Delta n_{(0,i]} \sim \textrm{Gamma}\left(a_\kappa+\sum_{l=1}^i\sum_{j=1}^m  \Delta n_{2,\tau_{l,j}},b_\kappa+\sum_{l=1}^i\sum_{j=1}^{m} g_2(x_{\tau_{l,j-1}})\Delta\tau\right),
\end{equation}
\begin{equation}\label{cp2}
\gamma|\Delta n_{(0,i]} \sim \textrm{Gamma}\left(a_\gamma+\sum_{l=1}^i\sum_{j=1}^m  \Delta n_{3,\tau_{l,j}},b_\gamma+\sum_{l=1}^i\sum_{j=1}^{m} g_3(x_{\tau_{l,j-1}})\Delta\tau\right),
\end{equation}
where $g_{2}(x)=h_2(x)/\kappa$ and $g_{3}(x)=h_{3}(x)/\gamma$, that is, the combinatorial factors in the hazard functions governing the infection and removal events. Similarly, if the discretised SDE (\ref{sde}) is adopted for the contact rate process $\tilde{\beta}_t$, then we may ascribe the prior $\lambda\sim \textrm{Gamma}(a_\lambda,b_\lambda)$ leading to the conditional posterior
\begin{equation}\label{cp3}
\lambda| \tilde{\beta}_{[0,i]} \sim \textrm{Gamma}\left(a_\lambda +\frac{i m}{2}, b_\lambda +\frac{1}{2}\sum_{l=1}^i\sum_{j=1}^{m}(\tilde{\beta}_{\tau_{l,j-1}}-\tilde{\beta}_{\tau_{l,j}})^2 /\Delta\tau \right).
\end{equation}
Hence, the conditional posterior $\pi(\theta,\lambda |\Delta n_{(0,i]}, \tilde{\beta}_{[0,i]})$ can be summarised by a vector of sufficient statistics $T_i:=T_i(\Delta n_{(0,i]},\tilde{\beta}_{[0,i]})$ given by the 3 shape and 3 rate hyper-parameters in (\ref{cp1})--(\ref{cp3}). This vector can then be updated recursively within the particle filter, upon initialising with $T_0=(a_\kappa,a_\gamma,a_\lambda,b_\kappa,b_\gamma,b_\lambda)$. In turn, particles $\theta^{(k)}$ and $\lambda^{(k)}$ can be updated by drawing from (\ref{cp1})--(\ref{cp3}) conditional on $T_i^{(k)}:=T_i\left(\Delta n_{(0,i]}^{(k)},\tilde{\beta}_{[0,i]}^{(k)}\right)$. 

Our treatment of $\phi$ depends on the observation model used. In the case of the Binomial model in (\ref{eqn:obseqn1}), with $\phi=\rho$, the conditional posterior $\pi(\phi|\Delta n_{(0,i]},y_{[1,i]})$ is tractable upon ascribing a $\textrm{Beta}(a_{\phi},b_{\phi})$ prior to $\phi$. We obtain
\[
\phi|\Delta n_{(0,i]},y_{[1,i]} \sim \textrm{Beta}\left( a_{\phi}+\sum_{l=1}^i y_{t_{l}},b_{\phi}+P' \sum_{l=1}^i \Delta N_{t_l} - \sum_{l=1}^i y_{t_l}\right).
\]

Unfortunately, in the case of the Negative Binomial model (\ref{eqn:obseqn2}) with $\phi=(\rho,\nu)'$, $\pi(\phi|\Delta n_{(0,i]},y_{[1,i]})$ remains intractable. We therefore treat $\phi$ as a dynamic parameter and induce artificial evolution of $\phi$ by following the approach of \cite{liu2001}. That is, if $\phi_i^{(k)}$ denotes the $k$th particle at time $t_i$, we propagate forward to time $t_{i+1}$ via 
\[
\phi_{i+1}^{(k)}\sim N\left(m_i^{(k)}, s^2 V_i \right)
\]
where $m_i^{(k)}=a\phi_i^{(k)}+(1-a)\bar{\phi}_i$, $V_i=\sum_{k=1}^N (\phi_i^{(k)}-\bar{\phi}_i)(\phi_i^{(k)}-\bar{\phi}_i)'/N$ and $\bar{\phi}_i=\sum_{k=1}^N \phi_i^{(k)}/N$. The practitioner can choose the hyper-parameters $a$ and $s$ to control shrinkage and over-dispersion. We refer the reader to \cite{liu2001} for typical choices of these hyper-parameters. 

We propagate the dynamic $\tilde{\beta}$ process blindly by drawing $\tilde{\beta}_{(i,i+1]}^{(k)}\sim \pi(\cdot|\tilde{\beta}_{t_i}^{(k)},\lambda^{(k)})$. To mitigate against jump process trajectories being inconsistent with the next observation, we propagate  $\Delta n_{(i,i+1]}^{(k)}$ conditional on $y_{t_{i+1}}$, by drawing 
\[
\Delta n_{(i,i+1]}^{(k)}\sim q\left(\cdot|\tilde{\beta}_{[i,i+1)}^{(k)},\phi_{i+1}^{(k)},\theta^{(k)},y_{t_{i+1}}\right)
\]
for some suitable construct $q\left(\cdot|\tilde{\beta}_{[i,i+1)}^{(k)},\phi_{i+1}^{(k)},\theta^{(k)},y_{t_{i+1}}\right)$, discussion of which is deferred until Section~\ref{sec:guide}. The complete particle filtering scheme (as appropriate for observation model (\ref{eqn:obseqn2})) is summarised in Algorithm~\ref{alg:PF}.

\begin{algorithm}[t]
        \caption{Particle Filter}
        \label{alg:PF}
        \textbf{Input:} data $y=(y_{t_1},\ldots, y_{t_L})'$, number of particles $N$, initial draws $\psi^{(1:N)}\sim \pi(\psi)$, $x_{t_0}^{(1:N)}\sim \pi(x_{t_0})$, $\tilde{\beta}_{t_0}^{(1:N)}\sim \pi(\tilde{\beta}_{t_0})$ and sufficient statistic $T_0^{1:N}$.\\
For $i=0,\ldots, L-1$ and $k=1,\ldots,N$:
\begin{enumerate}
    \item \textbf{Propagate} dynamic processes:
    \begin{itemize}
        \item[a.] draw $\phi_{i+1}^{(k)}\sim N(m_i^{(k)}, s^2 V_i )$,
        \item[b.] draw $\tilde{\beta}_{(i,i+1]}^{(k)}\sim \pi(\cdot|\tilde{\beta}_{t_i}^{(k)},\lambda^{(k)})$,
        \item[c.] draw $\Delta n_{(i,i+1]}^{(k)}\sim q(\cdot|\tilde{\beta}_{[i,i+1)}^{(k)},\phi_{i+1}^{(k)},\theta^{(k)},y_{t_{i+1}})$.
    \end{itemize} 
    \item \textbf{Resample} with replacement among $\left(\psi^{(1:N)},\tilde{\beta}_{t_{i+1}}^{(1:N)},\Delta n_{t_{i+1}}^{(1:N)}\right)$ using the weights
    \[
     w_{i+1}^{(k)} \propto 
     \frac{\pi(\Delta n_{(i,i+1]}^{(k)}|\tilde{\beta}_{[i,i+1)}^{(k)},\theta^{(k)})
     \pi(y_{t_{i+1}}|\Delta n_{t_{i+1}}^{(k)},\phi^{(k)})}
     {q(\Delta n_{(i,i+1]}^{(k)}|\tilde{\beta}_{[i,i+1)}^{(k)},\phi_{i+1}^{(k)},\theta^{(k)},y_{t_{i+1}})}
    \]
     as probabilities.
    \item \textbf{Update} sufficient statistic $T_{i+1}^{(k)}:=T_{i+1}(T_i^{(k)}, \Delta n_{(i,i+1]}^{(k)},\tilde{\beta}_{(i,i+1]}^{(k)})$. 
    \item \textbf{Sample} $\theta^{(k)}\sim \pi(\theta|T_{i+1}^{(k)})$ and $\lambda^{(k)}\sim \pi(\lambda|T_{i+1}^{(k)})$ using (\ref{cp1})--(\ref{cp3}).
\end{enumerate}

\textbf{Output:} particle representation  $\left(\psi^{(1:N)}_{i},\tilde{\beta}_{t_i}^{(1:N)},\Delta n_{t_i}^{(1:N)}\right)$ of the filtering distributions $\pi(\psi,\tilde{\beta}_{t_i},\Delta n_{t_i}|y_{[1,i]})$, $i=1,\ldots,L$.     
\end{algorithm}

\subsection{Bridge construct}\label{sec:guide}

As previously discussed, we wish to propagate particle trajectories conditional on the next observation. To this end, we derive an approximate instantaneous rate or hazard function that is conditioned on the next observation. The derivation involves the construction of Gaussian approximation to the joint distribution of the incidence over a time window whose right end-point is the next observation time, and the next observation itself. The conditioned hazard is then taken to be proportional to the expectation of the incidence given the observation. 

Suppose we receive an observation $y_{t_i}$ and have simulated as far as $\tau_{i,j} \in (t_{i-1},t_i]$ so that $x_{\tau_{i,j}}$ and $\Delta n_{(t_{i-1},\tau_{i,j}]}$ are fixed and known. By analogy with the unconditioned hazard function, we denote the conditioned hazard function by $h^*\left(x_{\tau_{i,j}}|y_{t_i}\right)$, with dependence on the parameters $\theta$ and incidence process suppressed for notational simplicity. 

Assuming a Normal approximation to the Poisson distribution, the number of reaction events in the interval $(\tau_{i,j},t_i]$, denoted by $\Delta N_{(\tau_{i,j},t_i]}$, is 
\begin{equation}
\label{eqn:Napprox}
\Delta N_{(\tau_{i,j},t_i]} \overset{\text{approx.}}{\sim} \textrm{N} \left\{h(x_{\tau_{i,j}})(t_i-\tau_{i,j}),H(x_{\tau_{i,j}})(t_i-\tau_{i,j})\right\}
\end{equation}
where $H(x_{\tau_{i,j}})=\textrm{diag}\left\{h(x_{\tau_{i,j}})\right\}$. Moreover, if we take a Normal approximation to the observation model, we obtain
\begin{align*}
Y_{t_i}|\Delta N_{(\tau_{i,j},t_i]} &\overset{\text{approx.}}{\sim}
\textrm{N}\left\{\mu(\Delta n_{(t_{i-1},\tau_{i,j}]}),
\sigma^2(\Delta n_{(t_{i-1},\tau_{i,j}]})
\right\}
\end{align*}
where 
$
\mu(\Delta n_{(t_{i-1},\tau_{i,j}]}) = \rho P'  (\Delta n_{(t_{i-1},\tau_{i,j}]} + \Delta N_{(\tau_{i,j},t_i]} ) 
$
for both Binomial and Negative Binomial observation models, with dependence on $\Delta N_{(\tau_{i,j},t_i]}$ suppressed for notational ease. The variance term is model dependent, taking the form
\[
\sigma^2(\Delta n_{(t_{i-1},\tau_{i,j}]}) = \rho(1-\rho)P'( \Delta n_{(t_{i-1},\tau_{i,j}]} + \Delta \hat{N}_{(\tau_{i,j},t_i]} ) 
\]
in the case of a Binomial observation model and 
\[
\sigma^2(\Delta n_{(t_{i-1},\tau_{i,j}]}) = \hat{\mu}(\Delta n_{(t_{i-1},\tau_{i,j}]}) + \hat{\mu}(\Delta n_{(t_{i-1},\tau_{i,j}]})^2/\nu
\]
in the case of a Negative Binomial observation model. Note that $\hat{\mu}$ has $\Delta N_{(\tau_{i,j},t_i]}$ replaced by an estimate $\Delta \hat{N}_{(\tau_{i,j},t_i]}$, leading to a linear Gaussian structure, which is essential for the tractability of the conditioned hazard function. We take $\Delta \hat{N}_{(\tau_{i,j},t_i]}$ to be the mean of the approximating Normal distribution in (\ref{eqn:Napprox}).

Making use of the linear Gaussian structure, we now form the approximate joint distribution of $\Delta N_{(\tau_{i,j},t_i]}$ and $Y_{t_i}$. This is 
\begin{align*}
    \begin{pmatrix}
    \Delta N_{(\tau_{i,j},t_i]} \\
    Y_{t_i}
    \end{pmatrix}
    \sim
    N
    \Bigg\{
    &\begin{pmatrix}
    h(x_{\tau_{i,j}})(t_i-\tau_{i,j}) \\
    \hat{\mu}(\Delta n_{(t_{i-1},\tau_{i,j}})
    \end{pmatrix},\\
    &\left.\begin{pmatrix}
    H(x_{\tau_{i,j}})(t_i-\tau_{i,j}) & 
    \rho H(x_{\tau_{i,j}})P(t_i-\tau_{i,j}) \\
    \rho P' H(x_{\tau_{i,j}})(t_i-\tau_{i,j}) &
     \rho^2P'H(x_{\tau_{i,j}})P(t_i-\tau_{i,j}) + \sigma^2(\Delta n_{(t_{i-1},\tau_{i,j}]})
    \end{pmatrix}
    \right\}.
\end{align*}
Now, we condition on $Y_{t_i} = y_{t_i}$, take the expectation of the resulting distribution and divide by $(t_i - \tau_{i,j})$ to obtain the conditioned hazard function as $h^*(x_{\tau_{i,j}},\theta|y_{t_i})$
\begin{align*}
h^*(x_{\tau_{i,j}}|y_{t_i}) = h(x_{\tau_{i,j}}) + \rho P' H(x_{\tau_{i,j}})&(\rho^2P'H(x_{\tau_{i,j}})P(t_i-\tau_{i,j})+\sigma^2)^{-1} \\
&\times (y_{t_i} - \hat{\mu}(\Delta n_{(t_{i-1},\tau_{i,j}]})).
\end{align*}
It is then straightforward to generate end-point conditioned trajectories, for example, over an interval $(t_{i-1},t_i]$ by executing Algorithm~\ref{alg:PoL}, with $h$ replaced by $h^*$ and time step $\Delta\tau$. Moreover, the corresponding likelihood $q(\cdot|\tilde{\beta}_{[i-1,i)},\phi_{i},\theta,y_{t_{i}})$ is simply a product of Poisson probability mass functions, each with rate $h^*(x_{\tau_{i,j}}|y_{t_i})\Delta\tau$, for $j=0,\ldots,m-1$.

\section{Applications}
\label{sec:applications}

Here, we consider three different applications of the particle filter (Algorithm~\ref{alg:PF}) using both synthetic and real data to demonstrate the effectiveness of the scheme. The first application uses synthetic data generated using the SIR model. We compare the accuracy of the resulting marginal parameter posteriors with a pseudo-marginal Metropolis-Hastings scheme. The other two applications are real data examples, taken from \cite{fintzi2021linear} and \cite{aspannaus2021bayesian}, respectively. In all cases, we set the time-step of the dS(E)IR model to be $\Delta\tau=0.1$, which gave a reasonable balance between accuracy and computational cost. All algorithms were coded in R and run on a PC with a 2.6 GHz clock speed across 20 cores.  Computer code to reproduce the following experiments can be downloaded from https://github.com/Sam-Whitaker/Sequential\_Bayes\_Epi.

\subsection{Simulation study} \label{subsec:sim_study}

\begin{figure}[b]
    \centering
    \includegraphics{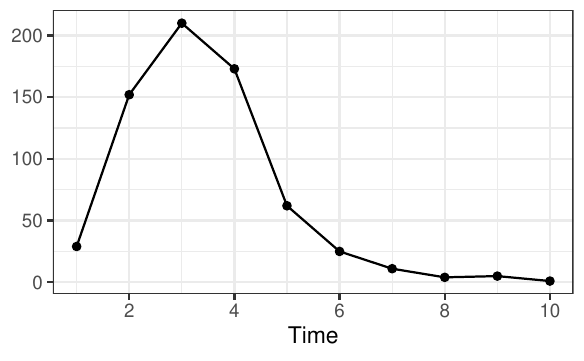}
    \caption{Synthetic data application. Number of new infections in $(t-1,t]$ for $t=1,\ldots,10$.}
    \label{fig:sim_data}
\end{figure}

We consider the dSIR model and synthetic data consisting of the (noisy) number of new infections in $(t-1,t]$ for $t=1,\ldots,10$. The data were generated by applying Algorithm~\ref{alg:PoL} with a time step of $0.001$ over the time interval $[0,10]$ and summing the number of infections over equally spaced sub-intervals of unit length. To emulate the influenza data set described in \cite{bmj1978}, we took the initial state to be $(i_0,s_0)' = (762,5)'$, a removal rate of $\gamma=0.5$ and assumed a time-varying infection rate whose logarithm is of the form
\[
\text{d}\tilde{\beta}_t = \lambda^{-1/2}\text{d}W_t, \quad \tilde{\beta}_0=-6 \quad \text{and} \quad \lambda=100
\]
so that the infection rate itself is scaled Brownian motion. Finally, we corrupted the data via a Binomial observation model with parameter $\rho=0.9$. The data are shown in \cref{fig:sim_data}.

\begin{figure}[t]
    \centering
    \includegraphics{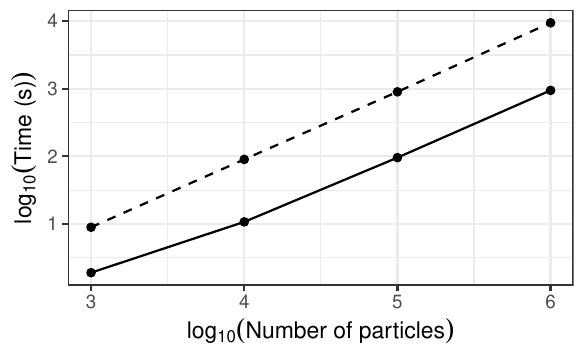}
    \caption{Synthetic data application. Wall clock CPU times (in seconds) for runs of the scheme in serial (dashed) and parallel (solid) versus the number of particles.}
    \label{fig:sim_timings}
\end{figure}

We adopted a weakly informative prior specification by taking $\log(\beta_0) \sim \text{N}(-6.5,0.5^2)$, $\gamma \sim \text{Gamma}(11,20)$, $\lambda \sim \text{Gamma}(15,0.14)$ and $\rho \sim \text{Beta}(90,15)$. The particle filter was run with $N=10^k$ particles, for $=1,2,\ldots,6$. The wall clock CPU time versus $N$ is shown in \cref{fig:sim_timings} for serial and parallel implementations; a single run with $N=10^6$ takes approximately 16 minutes. The benefit of parallelising the particle filter is clear, with an order of magnitude speed-up (over a serial implementation) achieved for $N\geq 10^4$ particles.  

\begin{figure}[ht!]
    \centering
    \includegraphics{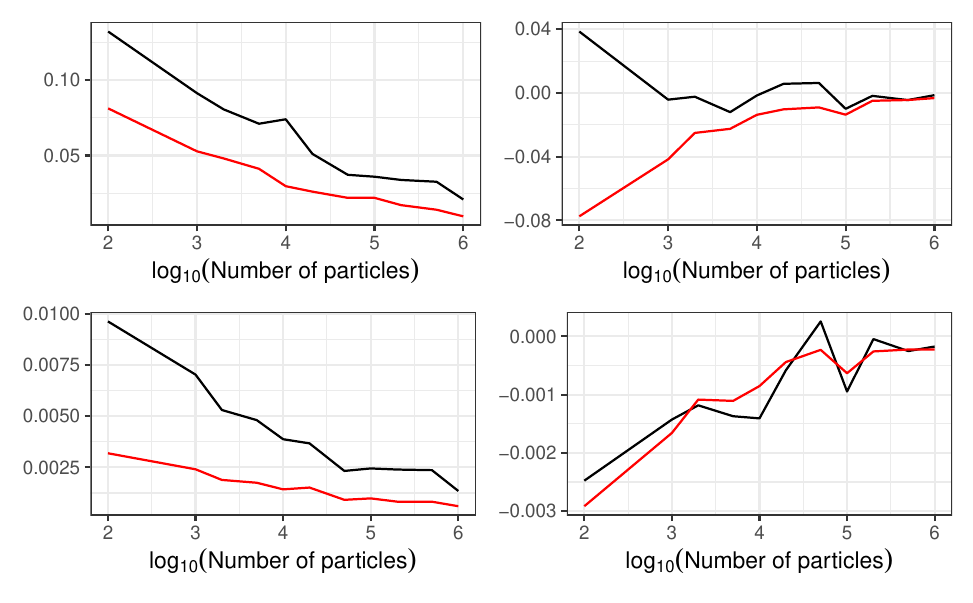}
    \caption{Synthetic data application. RMSE (left) and bias (right) versus $\log_{10}(N)$ for the particle filter's estimator of $e_1=\textrm{E}(\gamma|y)$ (black, top row) and $s_1=\textrm{SD}(\gamma|y)$ (red, top row) and $e_3=\textrm{E}(\rho|y)$ (black, bottom row) and $s_1=\textrm{SD}(\rho|y)$ (red, bottom row).}
    \label{fig:sim_rmse}
\end{figure}

\begin{figure}[ht!]
    \centering
    \includegraphics[]{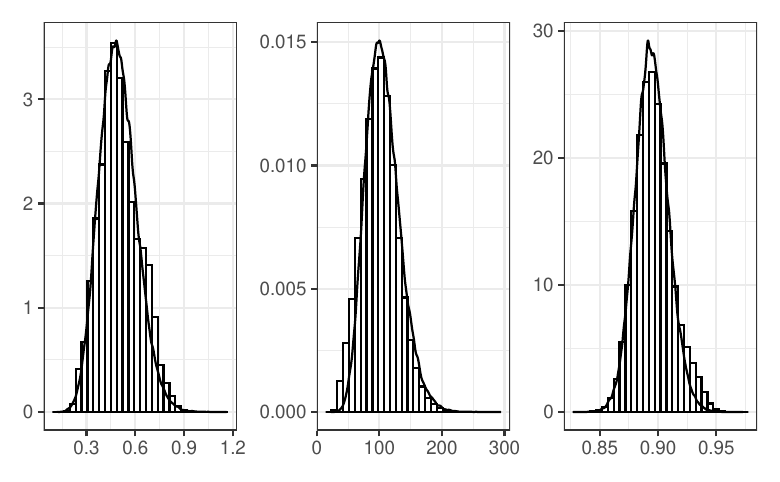}
    \caption{Synthetic data application. Posterior output from the particle filter (histograms) and ground truth posterior densities (kernel density estimates overlaid) obtained via PMMH ($10^6$ iterations). Panels left to right are $\gamma$, $\lambda$ and $\rho$ respectively.}
    \label{fig:sim_post}
\end{figure}

To benchmark accuracy for different values of $N$, we generated samples from the marginal parameter posteriors via a long run ($10^6$ iterations) of a pseudo-marginal Metropolis-Hastings scheme \citep[PMMH, e.g.][]{andrieu2010particle} targeting the posterior under the dSIR model. For each component of the parameter vector $\psi=(\gamma,\lambda,\rho)'$ we computed ``gold standard'' estimates of the marginal posterior expectations $\textrm{E}(\psi_i|y)$ and standard deviations $\textrm{SD}(\psi_i|y)$, denoted by $e_i$ and $s_i$, respectively. Replicate runs of the particle filter are then used to obtain corresponding estimates $\hat{e}_i^{(N,j)}$ and $\hat{s}_i^{(N,j)}$ where $N$ denotes the number of particles used and $j=1,\ldots,R$ indexes the replicate run. Bias and root mean squared error (RMSE) of the particle filter's estimator of $e_i$ is then calculated as 
\[
\textrm{Bias}(\hat{e}_i^{(N)})=\frac{1}{R}\sum_{j=1}^R (\hat{e}_i^{(N,j)}-e_i)
\quad \text{and} \quad
\textrm{RMSE}(\hat{e}_i^{(N)})= \sqrt{\frac{1}{R}\sum_{i=1}^R (\hat{e}_i^{(N,j)}-e_i)^2}
\]
with similar expressions obtained for the estimator of $s_i$. Bias and RMSE for the estimator of the marginal posterior expectation and standard deviation of $\gamma$ (removal rate) and $\rho$ (reporting rate) can be found in Figure~\ref{fig:sim_rmse}. 
Unsurprisingly, Bias and RMSE reduce as the number of particles increases. There is relatively little improvement in accuracy beyond $N=50K$, suggesting that for this scenario, this value of $N$ gives a reasonable balance between accuracy and efficiency. This remark is further supported by Figure~\ref{fig:sim_post}, which shows the particle filter sample output (after assimilating all observations) for $N=50K$, with the gold standard PMMH marginal posterior densities overlaid.   

\begin{figure}[ht!]
    \centering
    \includegraphics{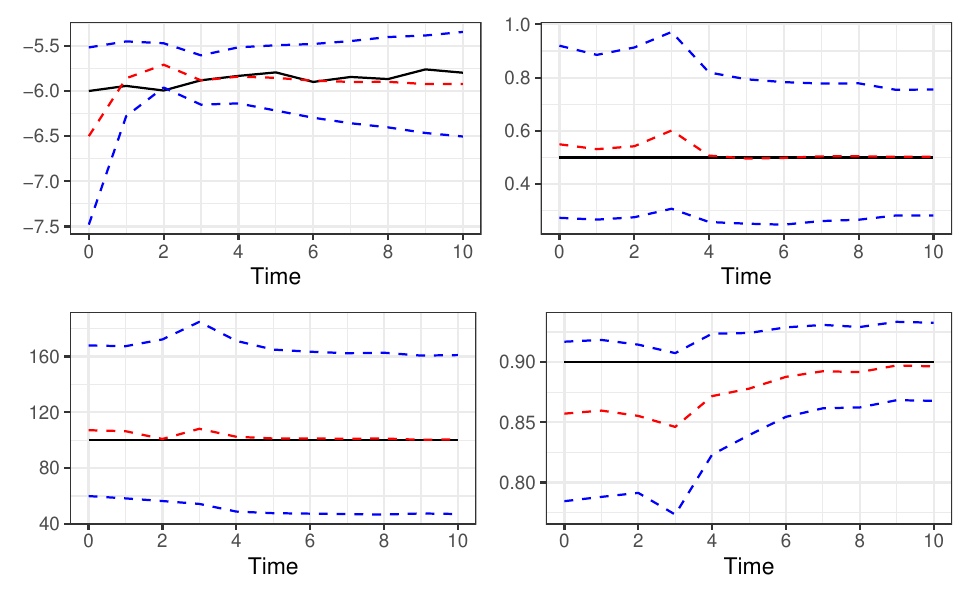}
    \caption{Synthetic data application. Filtering mean (red) and 95\% interval (blue) of the latent infection rate (top left), removal rate (top right), infection precision (bottom left) and reporting rate (bottom right). The ground truth is indicated (black).}
    \label{fig:sim_paras}
\end{figure}

Summaries (mean and 95\% credible interval) of the marginal filtering distributions $\pi(\psi_i|y_{[1,t]})$ and 
$\pi(\beta_t|y_[1,t])$ are shown in \cref{fig:sim_paras}. 
Samples from the filtering distributions are consistent with the ground truth parameter values that produced the data. Posterior uncertainty generally reduces as more data points are assimilated. However, there is little difference between the prior distribution of the precision parameter $\lambda$ governing the time-varying infection rate and the filtering distribution of $\lambda$ at each observation time-point; we anticipate that this parameter will be particularly sensitive to the choice of prior.   

\begin{figure}[ht!]
    \centering
    \includegraphics[]{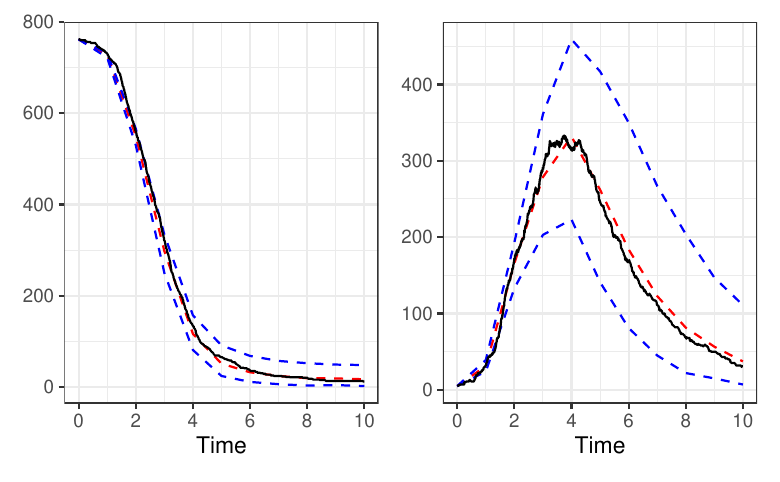}
    \caption{Synthetic data application. Filtering mean (red) and 95\% credible interval (blue) for the susceptible (left) and infective (right) species. The ground truth is shown in black.}
        \label{fig:sim_states}
\end{figure}

The particle filter gives samples of cumulative incidence over $(t_{i-1},t_i]$, $\Delta n_{t_i}^{(1:N)}$, from the (marginal) filtering distributions $\pi(\Delta n_{t_i}|y_{[1,i]})$, $i=1,\ldots,10$. Given corresponding samples of $x_{t_{i-1}}^{(1:N)}$, the prevalence at time $t_i$ can be computed for each sample $k$ via 
\[
x_{t_i}^{(k)} = x_{t_{i-1}}^{(k)} + \sum_j A_j' \Delta n_{j,t_i}^{(k)}.
\]
Hence, samples from the filtering distributions for prevalence, $\pi(x_{t_i}|y_{[1,i]})$, are easily generated as part of the particle filter. Summaries of these filtering distributions are shown in \cref{fig:sim_states}, from which we see that the filter output is consistent with the ground truth.

\subsection{Ebola}
\label{subsec:ebola}

\begin{figure}[ht!]
    \centering
    \includegraphics{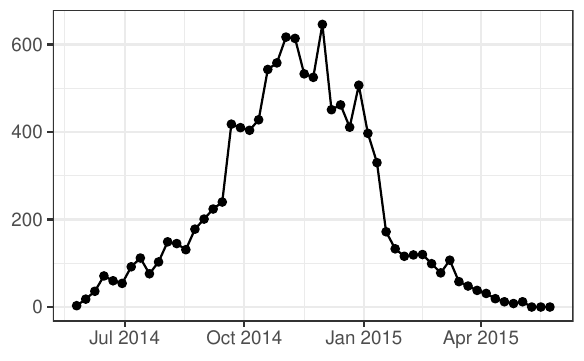}
    \caption{Ebola application. Weekly incidence data from the Ebola outbreak in Sierra Leone.}
    \label{fig:ebola_data}
\end{figure}

We consider the first of two real data examples by applying the proposed dSEIR model and inference scheme to a subset of the data found in \cite{fintzi2021linear}. In particular, we only consider data from Sierra Leone. The data are 53 weekly observations of the incidence of Ebola from May 2014 to May 2015; these are shown in \cref{fig:ebola_data}. 

Following \cite{fintzi2021linear}, we specify a Negative Binomial observation model to link the true unobserved incidence process to observations; see \cref{eqn:obseqn2}. The authors include a constant immigration of cases, occurring with rate $\alpha$, via a modification of the hazard of the first reaction, $h_1(x_t)$. 
As we are only considering one country in isolation, we ignore this proposed immigration, assuming that the import and export rates will be approximately equal. Instead, we fit the dSEIR model with a constant contact rate as described in \cref{sec:sem}.

Our choice of prior is consistent with the specification used in\cite{fintzi2021linear}; we take $\beta \sim \text{Gamma}(2,50000)$, $\kappa \sim \text{Gamma}(5,4.6)$, $\gamma \sim \text{Gamma}(10,10)$, $\text{logit}(\rho) \sim \text{N}(0.85,0.75^2)$ and $\nu \sim \text{Gamma}(5,0.2)$. We assume the initial state is fixed and known to be $x_0 = (44326,15,10)'$. We found that taking the number of particles to be at least $N=4\times 10^6$ is necessary to avoid degeneracy in the parameter samples governing the observation model (for which we resort to the jittering approach described in Section~\ref{sec:pf}. A single run of \cref{alg:PF} (parallelised over 20 cores) with this choice of $N$ took approximately 321 minutes (or around 5.3 hours). 

\begin{figure}[ht!]
    \centering
    \includegraphics{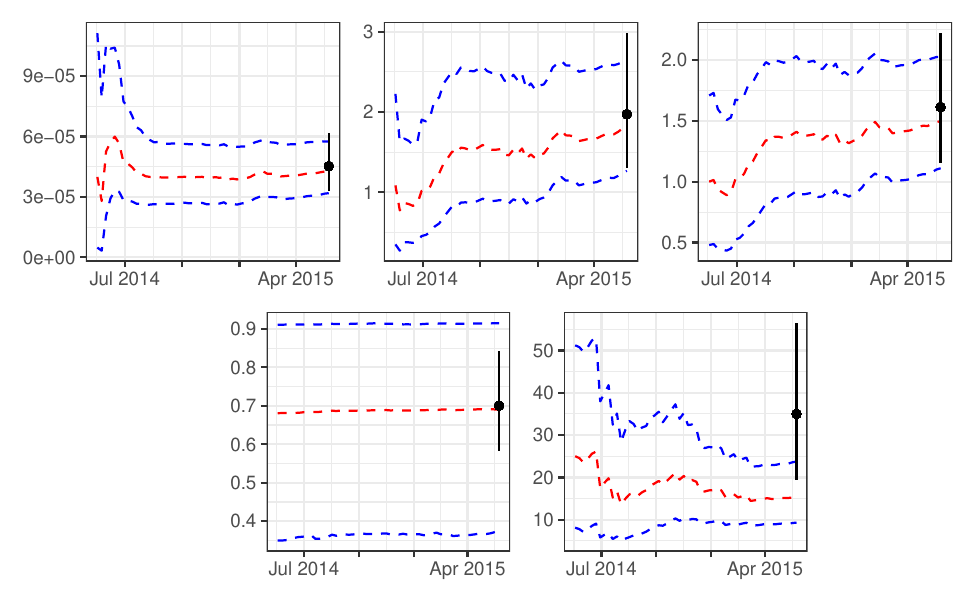}
    \caption{Ebola application. Filtering mean (red) and 95\% credible interval (blue) of the contact rate (top left), infection rate (top centre), removal rate (top right), reporting rate (bottom left) and overdispersion parameter (bottom right) for the dSEIR model. Overlaid are filtering means (black dots) and 95\% credible intervals (black lines) at the final time, using the linear noise approximation (LNA) as the transmission model.}
    \label{fig:ebola_paras}
\end{figure}

As in \cref{subsec:sim_study}, we provide summaries (mean and $95\%$ credible interval) of the marginal filtering distributions of each parameter in \cref{fig:ebola_paras}. We see that the uncertainty reduces from prior to posterior for most parameters, with the exception of the reporting rate $\rho$, which remains almost unchanged. This reduction in uncertainty is most prominent for both the contact rate, $\beta$ and the over-dispersion parameter $\nu$. 

\begin{figure}[ht!]
    \centering
    \includegraphics[]{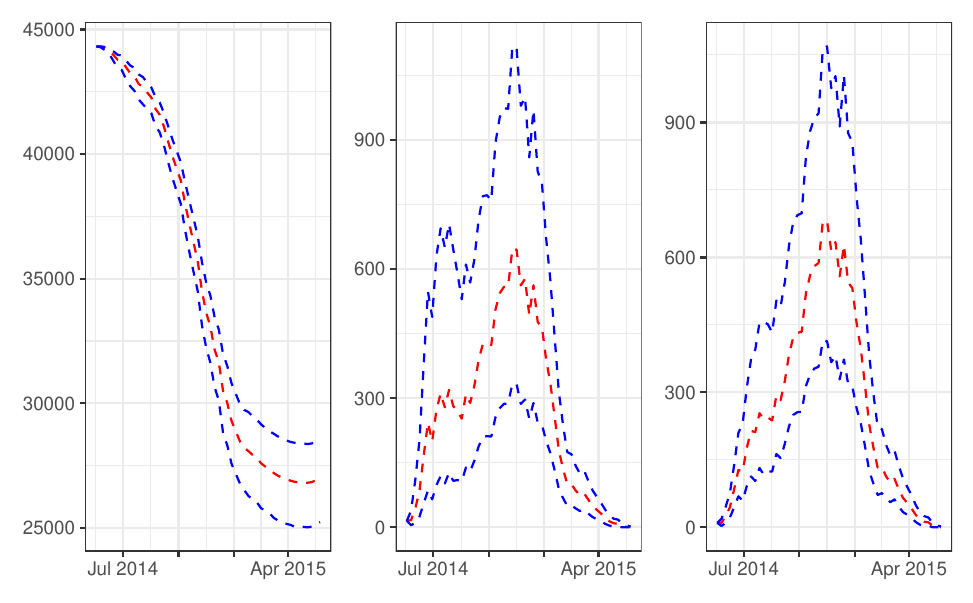}
    \caption{Ebola application. Filtering mean (red) and 95\% credible interval (blue) for the susceptible, exposed and infective species under the dSEIR model.}
    \label{fig:ebola_state}
\end{figure}

Given in \cref{fig:ebola_state} are summaries of the filtering distributions for the prevalence. We see that the credible region for the susceptible species narrows towards the middle of the data (around Nov-Dec 2014) before widening towards the end again. Conversely, the uncertainty increases towards the middle of the data in both the exposed and infectious species. The peaks in both the exposed and infectious species correspond to peaks in the data, which is to be expected; as the population of the exposed/infectious species increases, so will the weekly incidence.

\begin{figure}[ht!]
    \centering
    \includegraphics[]{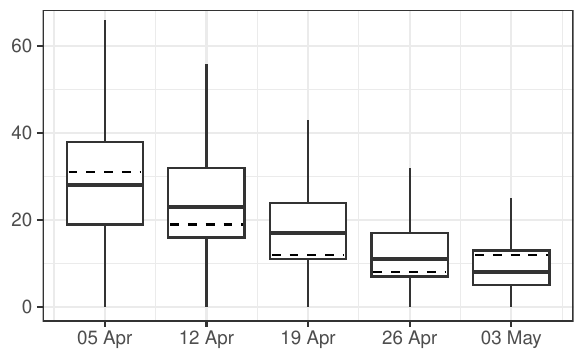}
    \caption{Ebola application. Five one step ahead dSEIR forecasts of the final five (non-zero) observed incidences overlaid with the true observed values (dashed).}
    \label{fig:ebola_pf_forecast}
\end{figure}

Finally, we produce five one-step-ahead forecasts of the weekly incidence for the last five non-zero observations in the data. To produce a forecast for the observed weekly incidence at an arbitrary time $T$, we run \cref{alg:PF} up to time $T-1$, and then, using the particle representation of $\pi(\phi,\tilde{\beta}_{[0,T-1]},\Delta n_{(0,T-1]}|y_{[1,T-1]})$ from the output of \cref{alg:PF}, we generate $N$ realisations of the cumulative incidence process over $(T-1,T]$. The final cumulative incidences are subjected to noise as per the observation model, and the resulting samples are then summarised using a box and whisker plot. We repeat this process for each of five forecast intervals of interest. The resulting box and whisker plots are provided in \cref{fig:ebola_pf_forecast}. We see that the observed values, given as dashed lines, lie close to the median value for each of the five forecasts and fall between the lower and upper quartiles, suggesting that the dSEIR model (in conjunction with the proposed inference scheme) is able to accurately forecast Ebola incidence.  

\subsubsection{Comparison with a linear noise approximation (LNA)}

Here, we compare forecast accuracy of the dSEIR model to the approach taken by \cite{fintzi2021linear}, who describe SEIR dynamics using a linear noise approximation \citep[see e.g.][]{Komorowski09,fearnhead12,schnoerr17}. That is, the LNA is used as the transmission model and the previously described Negative Binomial model is used as the observation model. In order to generate samples from the marginal parameter posterior, an MCMC scheme is used. In particular, and for reasons of computational efficiency, we implement the marginal Metro`polis-Hastings scheme of \cite{golightly23}; see \cref{app:LNA} for brief details regarding the LNA and the MCMC scheme used in this setting.   


The LNA requires solution of a coupled ordinary differential equation (ODE) system. We used a simple Euler solver with a time step of 0.01, which gave a reasonable computational efficiency and accuracy balance. The prior specification is as previously described. We ran the MCMC scheme for 100k iterations, using a correlated random walk proposal, tuned using the posterior variance from a shorter pilot run of 10K iterations. The resulting computational cost is approximately equal (in terms of CPU wall clock time) to the particle filter run with $N=4\times 10^6$.

Posterior means and 95\% credible intervals for the model parameters under the LNA are compared to the same posterior summaries obtained from the particle filter applied to the dSEIR model in \cref{fig:ebola_paras}. It is evident that posterior summaries under the LNA are mostly consistent with those obtained under the dSEIR model, with the exception of the overdispersion parameter, $\nu$, whose posterior mass is concentrated around larger values than that from the particle filter. There is a clear reduction in reporting rate posterior variance under the LNA compared to dSEIR, although the mean from both approaches appears consistent.

\begin{figure}
    \centering
    \includegraphics{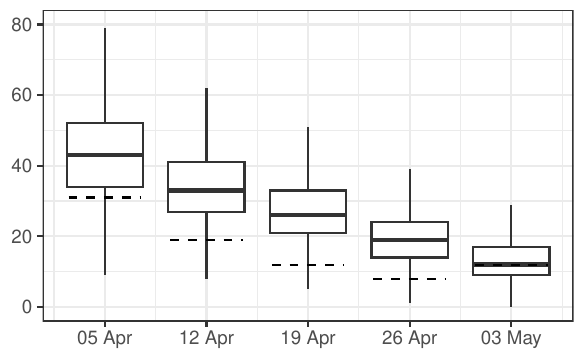}
    \caption{Ebola application. Five one step ahead LNA forecasts of the final five (non-zero) observed incidences overlaid with the true observed values (dashed).}
    \label{fig:ebola_LNA_forecast}
\end{figure}

As a final comparison, we now generate five one-step-ahead forecasts of the last five non-zero observations under the LNA. To generate a forecast at time $T$, we ran the LNA MCMC scheme using all the data up to and including time $T-1$, storing each accepted parameter vector along with a corresponding draw from $N_{T-1}|Y_{1:T-1}$. We then propagate a thinned sample forward over the interval $(T-1,T)$ to obtain estimates of the cumulative incidence at time $T$, which we subjected to noise according to the observation model. The forecasts are summarised by box and whisker plots in \cref{fig:ebola_LNA_forecast}.
A comparison of Figures~\ref{fig:ebola_pf_forecast} and \ref{fig:ebola_LNA_forecast} suggests that the LNA tends to overestimate the observation, with four out of five median forecasts lying below the lower quartile.

\subsection{Covid-19 in New York}
\label{subsec:covid}

\begin{figure}[ht!]
    \centering
    \includegraphics{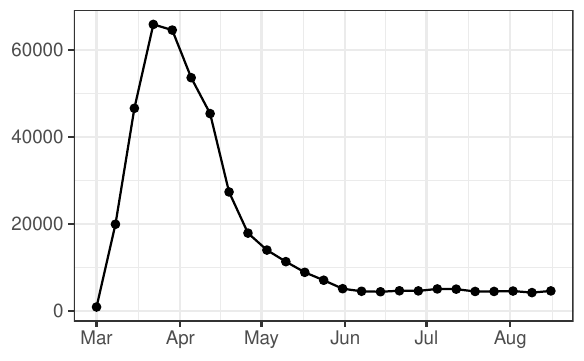}
    \caption{COVID-19 application. Weekly incidence in New York.}
    \label{fig:covid_data}
\end{figure}

For the final real data application, we consider a data set taken from \cite{aspannaus2021bayesian}. The data consist of twenty-five observations of the weekly incidence of COVID-19 in New York, starting in March 2020 and ending in mid-August 2020. The data are presented in \cref{fig:covid_data}. 

We adopt an approach consistent with the authors by modelling the data using an SIR model with dynamic contact and reporting rates, and a Negative Binomial observation model. Specifically, we allow the logit of the reporting rate to vary through time according to a scaled Brownian motion process whose precision is given by $\lambda_\rho$. To avoid any ambiguity, we denote the precision for the scaled Brownian motion process governing the log of the contact rate by $\lambda_\beta$. Hence, the dSIR transmission model is
\begin{equation}\label{dsir}
\left\{ \begin{array}{l}
X_{t+\Delta t}= x_{t} + A' \Delta N_{t},\\
\log \beta_{t+\Delta t}=\log \beta_t + \lambda_{\beta}^{-1/2} \Delta W_{1,t}, \\
\text{logit} \rho_{t+\Delta t}=\text{logit} \rho_t + \lambda_{\rho}^{-1/2} \Delta W_{2,t}.       
\end{array}
\right.
\end{equation}
where $W_{1,t}$ and $W_{2,t}$ are uncorrelated Brownian motion processes. The observation model is
\begin{equation}\label{COVIDobs}
Y_{t_i}|\Delta N_{t_i} \sim \textrm{NegBin}(\mu_i= \rho_{t_i} P'\Delta N_{t_i},\sigma^2_i=\mu_i+\mu_i^2/\nu), \quad i=1,\ldots, L. 
\end{equation}

We follow \cite{aspannaus2021bayesian} by adopting a prior specification as follows. We let $\log(\beta_0) = 0$, $\gamma \sim \text{Gamma}(11.088,2.192)$, $\lambda_{\beta} \sim \text{Gamma}(4,1)$, $\lambda_{\rho} \sim \text{Gamma}(4,1)$, $\rho_0 \sim \text{Beta}(3,2)$ and $1/\sqrt{\nu} \sim \text{U}(0,0.5)$ \emph{a priori}. We found that taking the number of particles to be at least $N=4\times 10^6$ avoids sample impoverishment. A single run of \cref{alg:PF} (parallelised over 20 cores) with this choice of $N$ took approximately 49 minutes.

\begin{figure}[ht!]
    \centering
    \includegraphics{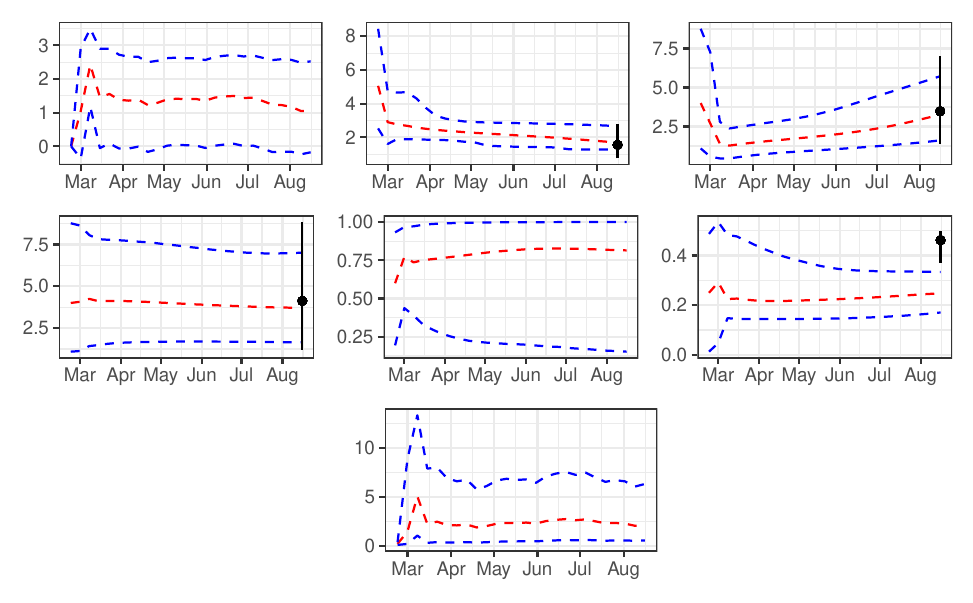}
    \caption{COVID-19 application. Mean (red) and 95\% interval (blue) from the filtering distributions of $\log(N\beta_t)$ (top left), $\gamma$ (top centre), $\lambda_\beta$ (top right), $\lambda_\rho$ (centre left), $\rho_t$ (centre), $1/\sqrt{\nu}$ (centre right) and the basic reproductive number $R_0$ (bottom) under the dSIR model. In black are the mean (dots) and 95\% credible intervals (lines) for each static parameter under the transmission model of Spannaus et al.}
    \label{fig:covid_para}
\end{figure}

\begin{figure}[ht!]
    \centering
    \includegraphics[]{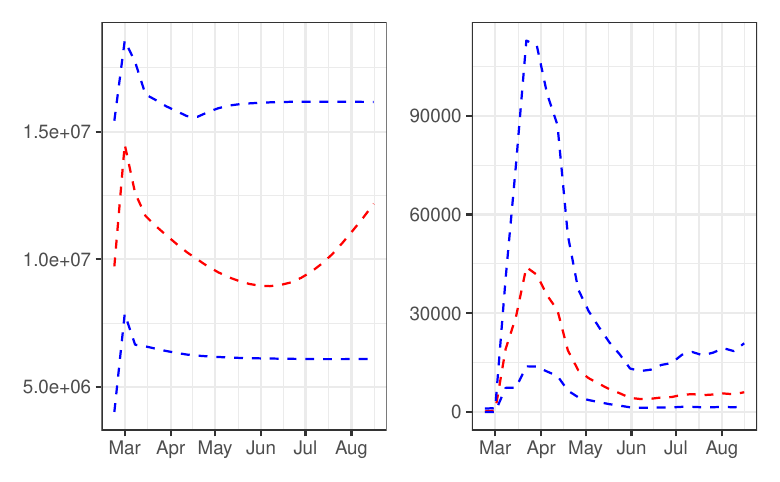}
    \caption{COVID-19 application. Mean (red) and 95\% interval (blue) for the filtering distributions of the Susceptible (left) and Infective (right) species.}
    \label{fig:covid_state}
\end{figure}

Summaries of the filtering distributions of the parameters and prevalence process are given in \cref{fig:covid_para,fig:covid_state}. We can see that the uncertainty about the parameters reduces from prior to posterior in all cases, indicating that the analysis is informative. 


Finally, we assess forecast accuracy of the dSIR model by providing five one-step-ahead forecasts for the final five non-zero observations, using the Monte Carlo method described in \cref{subsec:ebola}. It is evident that all forecast medians are consistent with the observations, suggesting good forecast accuracy. The forecast for the final observation is comparable to that given in \cite{aspannaus2021bayesian} (see Figure~5 therein). Further comparisons with the transmission model used in the latter are considered in the next section.  

\begin{figure}[ht!]
    \centering
    \includegraphics[]{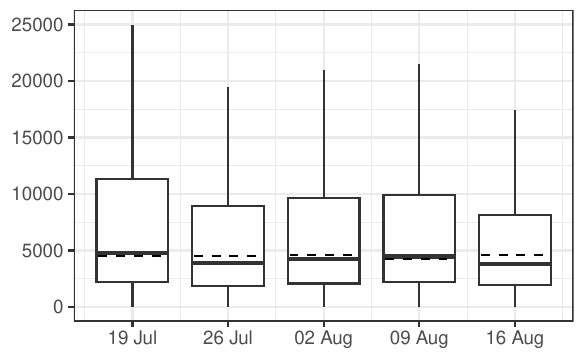}
    \caption{COVID-19 application. Five one-step-ahead dSIR forecasts of the final five observed incidences overlaid with the true observed values (dashed).}
    \label{fig:covid_forecast}
\end{figure}

\subsubsection{Comparison with Spannaus et al.}
\label{subsubsec:covidPMMH}
The SIR transmission model in \cite{aspannaus2021bayesian} takes the form of the well studied Kermack-McKendrick ordinary differential equation (ODE) system, albeit with a time-varying infection rate. That is
\begin{equation}\label{span}
\left\{ \begin{array}{l}
\text{d}S_t=-\beta_t S_tI_t \text{d}t,\\
\text{d}I_t=(\beta_t S_t I_t -\gamma I_t) \text{d}t,\\
\text{d}R_t = \gamma I_t \text{d}t,
\end{array}
\right.
\end{equation}
where $\log\beta_t$ follows an SDE with solution as in (\ref{dsir}). To facilitate comparison with the dSIR model, we assume that the logit reporting rate $\text{logit} \rho_t$ follows an SDE with solution as in (\ref{dsir}) and a Negative Binomial observation model as in (\ref{COVIDobs}).   

To generate samples from the marginal parameter posterior under the assumption of (\ref{span}), we ran the pseudo-marginal Metropolis-Hastings (PMMH) scheme described in \cite{aspannaus2021bayesian} for $10^6$ iterations. We report the posterior mean and 95\% credible interval for each of the static parameters in \cref{fig:covid_para}. It is clear that parameter samples are consistent with those obtained under the dSIR model (via the particle filter) for $\gamma$, $\lambda_\beta$ and $\lambda_\rho$. However, for the over-dispersion parameter $\nu$, there is some disagreement between the two marginal posterior distributions, with samples of $1/\sqrt{\nu}$ typically smaller under dSIR than when using the transmission model in (\ref{span}).

\begin{figure}[ht!]
    \centering
    \includegraphics[]{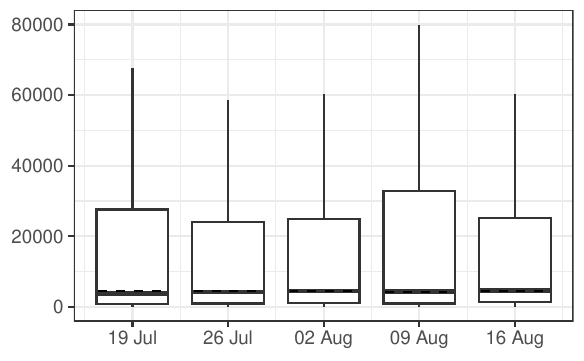}
    \caption{COVID-19 application. Five one-step-ahead forecasts of the final five observed incidences assuming the model of Spannaus et al, overlaid with the observed value (dashed).}
    \label{fig:covid_pmmh_forecast}
\end{figure}

We also provide five one-step-ahead forecasts under the assumption of the transmission model in (\ref{span}), for the final five non-zero observations in the data set in \cref{fig:covid_pmmh_forecast}. Comparing Figures~\ref{fig:covid_forecast} and \ref{fig:covid_pmmh_forecast}, we see agreement between forecast medians and the observations (for both schemes), although use of (\ref{span}) as the inferential and forecasting model appears to result in increased forecast uncertainty (relative to dSIR). This is likely due to the increased values of the inverse over-dispersion samples, in turn resulting in a larger observation variance.  

\section{Summary}
\label{sec:sum}

In this paper, we have considered the challenging problem of performing sequential Bayesian inference for the static parameters and latent, dynamic components of a stochastic representation of an epidemic compartment model. The inference task was developed under the assumption that only imperfect, discrete-time incidence values are available. 

The most natural continuous-time Markov jump process (MJP) representation of cumulative incidence was replaced by a discrete-time approximation in which the numbers of each transition event (e.g. exposure, infection or removal in an SEIR model) are assumed to be Poisson distributed over a time interval whose length is specified by the practitioner. The model allows for the straightforward incorporation of time varying parameters via time discretised stochastic differential equations (SDEs). The resulting dSEIR model admits several inferential advantages over other (stochastic) approximations to the MJP. Conditional on complete information regarding the cumulative incidence process, the static parameters governing the latent transmission model, and in some cases the observation model, follow a tractable posterior distribution. We leveraged this tractability via a sequential Monte Carlo scheme (particle filter) that rejuvenates parameter samples by drawing from their conditional parameter posterior after propagating and resampling particle trajectories of the latent incidence process. This approach essentially follows the framework of \cite{storvik2002} \citep[see also][for a related particle learning approach]{carvalho2010}. We further modified the particle filter by propagating particle trajectories conditional on the next observation, by adapting the conditioned hazard of \cite{golightly2015} to the cumulative incidence setting. 

We applied the resulting methodology to synthetic and real data examples, benchmarking against competing models and inference schemes. The proposed particle filter benefits from a parallel implementation, as demonstrated using synthetic data generated from a stochastic SIR model with a time-varying infection rate and Binomial observation model. We observed with an order of magnitude speed-up (over a serial implementation) for $N\geq 10^4$ particles. We also achieve satisfactory inferential accuracy, as measured by bias and RMSE for key posterior quantities, relative to corresponding values obtained from long runs of a pseudo-marginal Metropolis-Hastings scheme. We compared forecast accuracy of the dS(E)IR model to two competing approaches via real data applications involving the spread of Ebola in West Africa \citep{fintzi2021linear} and COVID-19 in New York \citep{aspannaus2021bayesian}. In the former, a linear noise approximation was used as a tractable approximation to the MJP representation of cumulative incidence. For a fixed computational budget, the dSEIR model better reflects the end of the epidemic, likely due to relatively low numbers of incidence, for which a continuous-valued approximation is likely to give a relatively poor approximation. For the final application involving COVID-19, we compared against a transmission model in which the infection rate of a deterministic ordinary differential equation (ODE) system is replaced with the solution of an SDE. We found that this approach leads to accurate point-summary forecasts but over-estimates uncertainty (relative to the dSIR model). It is also worth noting that our empirical findings suggest that use of the particle filter allows for assimilating a new observation in around 6 minutes for the dSEIR model and 2 minutes for the simpler dSIR model.   

\subsection{Limitations}
Here, we consider several limitations of the proposed approach and make some suggestions for future work. 

The dS(E)IR model requires specification of a step size, which balances accuracy (relative to the most natural MJP representation of cumulative incidence) and computational cost. We have assumed a fixed step size but note that an optimal choice is likely to depend on the phase of the epidemic. Conditions under which the Poisson approximation is reasonable have been discussed by \cite{gillespie2001} among others; for methods that allow a step size to be chosen adaptively, see e.g. \cite{cao2006efficient,Sandmann2009}.   

The particle filter used in this paper propagates particles conditional on the next observation by replacing the Poisson rate with a hazard function derived via a Gaussian approximation of the cumulative incidence between the current time and next observation, conditional on the observation itself. A further assumption in deriving the resulting conditioned hazard is linearity of the latent process over a time interval of length at most given by the observation interval. In scenarios in which the epidemic is sparsely observed in time, to the extent that cumulative incidence does not evolve in a linear fashion, the conditioned hazard is likely to result in trajectories that are inconsistent with the true conditioned process. Recent work on sampling conditioned jump processes (in a chemical kinetics setting) is relevant here \citep[see e.g.][]{golightlyMJP19,corstanje2023} and could be adapted to dS(E)IR model.     

A key feature of the proposed inference scheme is the leveraging of a tractable conditional parameter posterior, which depends (given complete information on the latent process up to time $t_i$) on a low-dimensional sufficient statistic $T_i$. This approach also forms the basis of particle learning \citep{carvalho2010}. Particle degeneracy in the latter suite of algorithms has been well documented \citep[see e.g. Section 4 of][]{chopinPL}. That is, the number of different values of the statistic at a fixed time point prior to $t_i$ that contribute to $T_i$ is decreasing in $i$ at an exponential rate. The applications presented here involve relatively short time-series coupled with a parallel implementation that allows sufficiently many particles to avoid degeneracy. Nevertheless, application to data sets involving longer time horizons remains of interest, as does a comparison with SMC schemes that use resample-move steps, such as SMC$^2$.

\clearpage

\bibliographystyle{apalike}
\bibliography{references}

\clearpage

\appendix

\section{LNA details for the Ebola application \label{app:LNA}}
In what follows, we give an intuitive derivation of the linear noise approximation of cumulative incidence under the assumption of an SEIR transmission model, followed by brief details of the inference scheme. We refer the reader to \cite{fintzi2021linear} \citep[see also][]{golightly23} for further details.

\subsection{LNA derivation \label{app:LNAder}}

Recall the three pseudo-reactions defining the SEIR compartment model:
\[
\mathcal{R}_1: S+I\overset{\beta}{\rightarrow}E+I, \qquad \mathcal{R}_2: E\overset{\kappa}{\rightarrow}I, \qquad \mathcal{R}_3: I\overset{\gamma}{\rightarrow}\emptyset.
\]
At any time $t$, the state of the prevalence process is given by the length three vector $x_t=(s_t,e_t,i_t)'$, and the state of the cumulative incidence process by $n_t=(n_{t,1},n_{t,2},n_{t,3})'$. Additionally, the hazard function is given by $h(x_t)=(\beta s_t i_t, \kappa e_t, \gamma i_t)'$. We know that the prevalence process $\{X_t, t\geq 0\}$ and the cumulative incidence process $\{N_t, t\geq 0\}$ are related via 
\[
X_t = x_0 + A' N_t
\]
where $x_0$ is the initial state and $A$ is the net effect matrix. It follows that, at any time $t$, the state of the prevalence process is given by
\begin{equation}
    \label{eqn:LNAprev1}
    x_t = (s_0 -n_{1,t}, e_0+n_{t,1}-n_{t,2}, i_0+n_{t,2}-n_{t,3})'.
\end{equation}
Furthermore, by substituting the relevant terms from (\ref{eqn:LNAprev1}) above into the expression for $h(x_t)$, we can re-write the hazard function in terms of $n_t$. The resulting expression, denoted by $h^*(n_t)$, is
\[
h^*(n_t) = (\beta[s_0-n_{t,1}][i_0+n_{t,2}-n_{t,3}],\kappa[e_0+n_{t,1}-n_{t,2}],\gamma[i_0+n_{t,2}-n_{t,3}])'.
\]
Since this example does not use a time-varying contact rate, the It\^{o} SDE which best matches the MJP is given by
\begin{equation}\label{cle2}
dN_t = h^*(n_t)dt + \text{diag}\left\{\sqrt{h^*(n_t)}\right\}dW_t
\end{equation}
resulting in drift and diffusion functions of $h(n_t)$ and $\text{diag}\left\{h(n_t)\right\}$ respectively. 

The SDE in (\ref{cle2}) can be linearised as follows. Consider a partition of $N_t$ as $N_t=\eta_t +R_t$ where 
$\eta_t$ is a deterministic process satisfying the ordinary differential equation (ODE)
\begin{equation}\label{lna1}
\frac{d\eta_t}{dt} = h^*(\eta_t)
\end{equation} 
and $R_t=N_t-\eta_t$ is a residual stochastic process satisfying a (typically) intractable SDE found by considering $dN_t-d\eta_t$. We obtain an approximate, tractable $\hat{R}_t$ by Taylor expanding $h^*(n_t)$ and $\operatorname{diag}\{h^*(n_t)\}$ about $\eta_t$. Retaining the first two terms in the 
expansion of the former and the first term in the expansion of the latter gives
\begin{equation}\label{approx_resid_sde}
d\hat{R}_t=F_t\hat{R}_t\,dt+ \operatorname{diag}\{\sqrt{h^*(\eta_t)}\}\,dW_t
\end{equation}
where $F_t$ is the Jacobian matrix with ($i$,$j$)th element given by the partial derivative of the $i$th component of $h^*(\eta_t)$ with respect to the $j$th component of $\eta_t$.
The resulting matrix is 
\[
F_t = 
\begin{pmatrix}
    \beta[-i_0-n_{t,2}+n_{t,3}] & \beta[s_0 - n_{t,1}] & \beta[-s_0 + n_{t,1}] \\
    \kappa & -\kappa & 0 \\
    0 & \gamma & -\gamma
\end{pmatrix}.
\]

For a fixed or Gaussian initial condition,  the solution of (\ref{approx_resid_sde}) is obtained as
\begin{equation}\label{lnaINC}
\left(N_t | N_{0}=\eta_0+r_0 \right) \sim \textrm{N}(\eta_t + G_t r_{0}, V_t)
\end{equation}
where the fundamental matrix $G_t$ satisfies
\begin{equation} \label{lna2}
\frac{dG_t}{dt} = F_t G_t, \quad G_{0} = I_3
\end{equation}
and $V_t$ satisfies
\begin{equation} \label{lna3}
\frac{dV_t}{dt} = V_t F_t' + \operatorname{diag}\{h^*(\eta_t)\} + F_t V_t, \quad V_0 = 0_{3}.
\end{equation}
Note that $I_3$ and $0_3$ are the $3\times 3$ identity and zero matrices respectively. The LNA for the cumulative 
incidence process is then summarised by (\ref{lnaINC}) and (\ref{lna1}), (\ref{lna2}), (\ref{lna3}).

\subsection{Bayesian inference using the LNA}

Let $n=\{n_t, t_0\leq t\leq t_L\}$ denote the latent (continuous-time) incidence process over $[t_0,t_L]$. The joint posterior density over the static parameters $\psi$ and latent incidence $n$ can be factorised as 
\begin{equation}\label{jpost2}
\pi(\psi, n|y) = \pi(\psi|y)\pi(n|\psi,y)
\end{equation}
where
\begin{equation}\label{jpost3}
\pi(\psi|y) \propto \pi(\psi)\pi(y|\psi).
\end{equation}
Sampling of (\ref{jpost2}) can be achieved via a two step approach in which samples are generated from the marginal parameter posterior $\pi(\psi|y)$, and then in the second step, a sample of the latent incidence process is generated from the conditional posterior $\pi(n|\psi,y)$ for each sample of $\psi$. As noted in \cite{golightly23}, both densities in the factorisation in (\ref{jpost2}) are tractable under the LNA, provided the construct linking the observations and latent transmission model is linear and Gaussian. 

For the Ebola data, we use a Negative Binomial model, whose form is given by (\ref{eqn:obseqn2}). To leverage the tractability of the LNA, we approximate the observation model via
\begin{equation}\label{eqn:obseqn3}
	Y_{t_i}|\Delta N_{t_i} \sim \textrm{N}(\rho P'\Delta N_{t_i},\hat{\mu}_i+\hat{\mu}_i^2/\nu)
\end{equation}
where $\hat{\mu}_i=\rho P'\textrm{E}(\Delta N_{t_i})$, and $\textrm{E}(\Delta N_{t_i})$ is the expected cumulative incidence over $(t_{i-1},t_i]$, under the LNA. Now, the resulting observed data likelihood can be efficiently evaluated by first noting the factorisation 
\begin{equation}\label{marglike}
\pi(y|\psi)=\pi(y_{t_0}|\psi)\prod_{i=1}^L \pi(y_{t_i}|y_{t_{0:(i-1)}},\psi)
\end{equation}
where $y_{t_{0:(i-1)}}=(y_{t_0},\ldots,y_{t_{i-1}})'$. The terms in (\ref{marglike}) can be recursively evaluated via a forward filter. Moreover, draws from the conditional posterior $\pi(n|\psi,y)$ can be obtained by backward sampling (using the output of the forward filter). For full details of the forward filter, backward sampling (FFBS) algorithm, see \cite{golightly23}.  







\end{document}